\documentclass[iop]{emulateapj}
\usepackage{color}
\usepackage{natbib}
\slugcomment{}

\shorttitle{Chemistry of distinct RGB groups in NGC~2808}
\shortauthors{Carretta}

\begin{document}
\title{Three discrete groups with homogeneous chemistry along the red giant 
branch in the globular cluster NGC~2808\altaffilmark{1}}

\author{Eugenio Carretta\altaffilmark{2}}

\altaffiltext{1}{Based on data collected at the ESO telescopes under 
programme 072.D-0507 and during the FLAMES Science Verification programme with
the UVES spectrograph.}
\altaffiltext{2}{INAF, Osservatorio Astronomico di Bologna, via Ranzani 1,
       40127,  Bologna,  Italy. eugenio.carretta@oabo.inaf.it}

\begin{abstract}

We present the homogeneous reanalysis of Mg and Al abundances from high
resolution UVES/FLAMES spectra for 31 red giants in the globular cluster
NGC~2808. We found a well defined Mg-Al anticorrelation reaching a regime of
subsolar Mg abundance ratios, with a spread of about 1.4 dex in
[Al/Fe]. The main result from the improved statistics of our sample is that the
distribution of stars is not continuous along the anticorrelation
as they are neatly clustered into three distinct clumps each
with different chemical composition. One group (P) shows the primordial 
composition of field stars of similar metallicity, and the other two (I and E) 
have increasing abundances of Al and decreasing abundances of Mg.  The fraction
of stars we found in the three components (P: 68\%, I: 19\%, E: 13\%) is in excellent
agreement with the number ratios computed for the three distinct main sequences
in NGC~2808: for the first time there is a clear correspondence between discrete
photometric sequences of dwarfs and  distinct groups of giants with homogeneous
chemistry.  The composition of the I group cannot be reproduced by mixing of 
matter with extreme processing in hot H-burning and gas with pristine,
unprocessed composition, as also found in the recent analysis of three discrete
groups in NGC~6752. This finding  suggests that different classes of polluters
were probably at work also in NGC~2808.
\end{abstract}


\section{Introduction}
Since decades the evidence of multiple stellar populations in globular clusters
(GCs) was hidden in plain sight in spectroscopic observations \citep{gra12}.
The large star-to-star abundance variations in light elements
(depletion of  C, O, Mg anticorrelated with enhancement of N, Na,
and Al) were understood as the simultaneous action of the NeNa and MgAl cycles
in the same regions where the ON part of the CNO cycle is fully operative in
H-burning at high temperature \citep{den89,lan93}.
The pivotal study by \citet{gra01} changed forever our
perspective. Their finding of Na-O and Al-Mg anticorrelations among unevolved
stars in NGC~6752 was the clearcut proof that these objects were formed by gas
polluted by ejecta of more massive stars of a past generation, since the
currently observed stars cannot synthesize the involved elements. Thus, whenever
such anticorrelations are traced, we are witnessing the presence of multiple
stellar generations. 

The most outstanding signature, the Na-O anticorrelation discovered by the
Lick-Texas group \citep[see][]{kra94}, has been the subject of our ongoing FLAMES
survey of 
24 GCs \citep[and references therein]{car06,car09a,car09b,car10,car14}. Well 
quantified by the interquartile range of the [O/Na]
ratio \citep[hereinafter Paper I]{carpap1} the homogeneous study of this feature
shows that the phenomenon of multiple populations is primarily driven by the
cluster total mass \citep{car10} and it is tightly linked to the horizontal
branch (HB) morphology \citep{car07a,gra10}. However, the Na-O
anticorrelation cannot provide useful enough information on the discrete or
continuous nature of multiple populations due to the intrinsic difficulty
in measuring oxygen abundances: often, only the broad division of stars in
first and second generations with different composition is revealed 
\citep[e.g.][]{mar08,car09a,joh12,cor14,car14}.

On the other hand, the increasing precision in photometric data detects split 
sequences at all evolutionary phases, mainly due to the work by Milone and
collaborators \citep[see][]{mil08,mil12,mil13}. In turn, discrete sequences mean
interruptions in the injection of ejecta in the intracluster gas, and by
calibrating the quiescent periods we may hope to provide more stringent
constraints on the still elusive nature of the early generation polluters such
as intermediate mass asymptotic giant branch (AGB) stars \citep{ven01} or
fast rotating massive stars \citep[FRMS;][]{dec07}. 

Very promising is the Mg-Al anticorrelation, only found in massive and/or
metal-poor clusters \citep[e.g.][]{car09b,yon05}. Recently \citet{car12} found
that in NGC~6752 red giant branch (RGB) stars are clearly clustered around three
distinct values of Al, with groups nicely corresponding to the three photometric
sequences revealed by Str\"omgren photometry on the RGB \citep{car11}. It seems
that the [Al/Mg] ratio is very efficient in enhancing the signal of homogeneous
subgroups, especially in GCs with [Fe/H]$\leq-1$. We then applied the same
approach to the globular cluster NGC~2808.

NGC~2808 is the ideal target for exploring the connection between distinct
photometric sequences and discrete populations with homogeneous chemistry. It is
massive \citep[$M_V=-9.39$ mag;][]{har96}, moderately metal-poor, with
three distinct main-sequences \citep[MS:][]{pio07,mil12} and a multimodal
distribution of stars on the HB \citep[see][]{bed00}. This
GC has been the target of extensive abundance analysis by our group. 
\citet{car04a} derived the Na-O anticorrelation in this cluster from 19 RGB stars,
while other light elements (including Mg, Al, Si) were studied for the same
sample in Paper I. Na, O abundances for 120 giants and Al, Mg abundances for
another limited sample of 12 RGB stars were obtained in \citet{car06} and
\citet{car09b}, respectively. Finally, \citet{gra11} analyzed high resolution
spectra of 42 stars on the HB of NGC~2808.

However, a complete study of the Mg-Al pattern in NGC~2808 was hampered by the
lack of large samples with homogeneous analysis, since the adopted scales of
atmospheric  parameters were different in Paper I and in \citet{car09b}.
In the present Letter we derive Mg and Al abundances for 31 red giants in
NGC~2808. The high degree of homogeneity of the analysis coupled with small
internal errors highlight that stars are clustered around three distinct levels
of Al and Mg. The number ratios of these subgroups are found to be in excellent
correspondence with the number of stars in the three  MSs.

\section{Data and analysis}

\begin{figure} 
\centering
\includegraphics[scale=0.80]{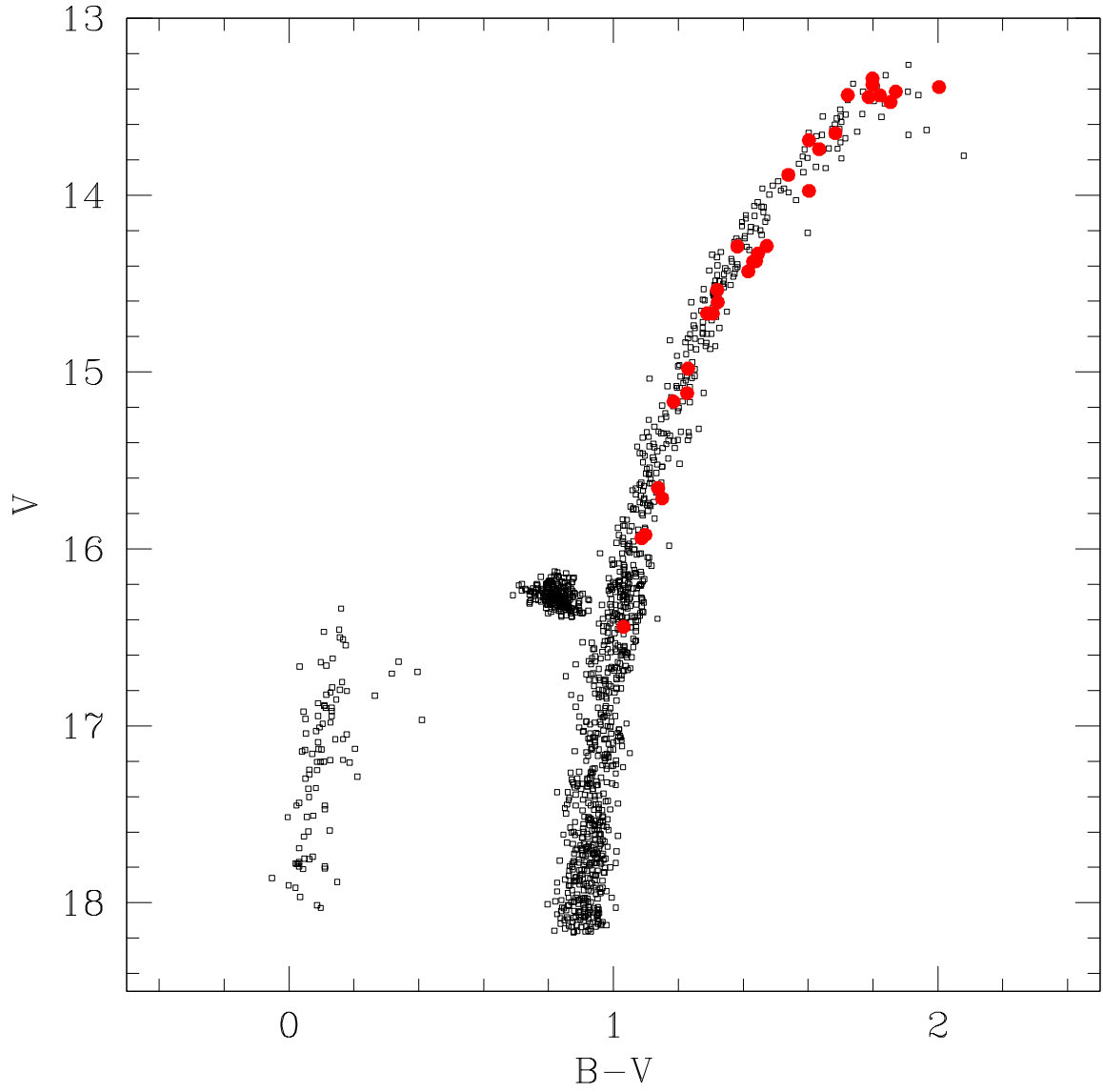}
\caption{$(V,B-V)$ color-magnitude diagram of NGC~2808 from Bedin et al. (2000).
Stars in the present study are marked with large filled circles.}
\label{f:fig1}
\end{figure}

Here we present the
abundances of Al and Mg for all the 31 stars observed with FLAMES/UVES. The UVES
Red Arm fed by FLAMES fibers provides a wavelength coverage from 4800 to
6800~\AA\ with a resolution of  $R\simeq 45,000$, hence abundances of
Al can be derived from the subordinate doublet at  6696-98~\AA\ (not available
in the GIRAFFE setups used for the Na-O anticorrelation).  The whole sample
include 19 stars from the FLAMES Science Verification (hereinafter sample SV,
see Paper I) and 12 giants from the FLAMES survey 
\citep[sample NAO,][ESO Programme  072.D-0507]{car06,car09b}, spanning the 
three brightest magnitudes along the RGB of NGC~2808 (see Fig.~\ref{f:fig1}).
Details of the observations {and data reduction} can be found in the original papers. 

The novelty of the present analysis is the use for the entire sample of our 
well tested procedure for the derivation of atmospheric parameters, in
particular the effective temperature T$_{\rm eff}$. First pass values are from
$V-K$ colors and the calibrations of \citet{alo99,alo01}. These were the
temperatures actually adopted in Paper I (sample SV). The final T$_{\rm eff}$
values for all targets were however obtained using  an average relation between
the temperatures derived in this first step and the apparent magnitude of
stars.  This procedure was successfully used in our survey of 24 GCs  to
decrease the star-to-star errors in abundances due to uncertainties in
temperatures, since magnitudes of bright RGB stars can be measured with higher
precision than colors. 

For NGC~2808, affected by relatively high differential \citep{bed00} reddening 
\citep[(E(B-V)=0.22 mag,][]{har96} the relation was derived as a function of the
$K$ magnitudes from 2MASS \citep{skr06}, minimizing the impact of the
differential reddening in the derived T$_{\rm eff}$. Optical $V$ magnitudes are
from  \citet{bed00}. For the 12 giants of the NAO sample the new temperatures
differ on average by about 5 K from the previous values. The new temperatures
are on average higher by 27 K, although with a large scatter (118 K), for the SV
sample.

Surface gravities were obtained from the position in the color-magnitude diagram
(CMD), using the derived T$_{\rm eff}$s, the distance modulus  $(m-M)_V=15.59$
mag from \citet{har96}, bolometric corrections from \citet{alo99}, a mass of
0.85~M$_\odot$\  for all stars and $M_{\rm bol,\odot} =4.75$ for the Sun, as in
our previous studies of other 23 GCs. Values of the microturbulent velocity
$v_t$ were obtained by eliminating trends in the relation between
abundances from Fe~{\sc i} lines and expected  line strength \citep{mag84}.

Equivalent widths ($EW$s) of the two Al lines and of two to three high
excitation Mg lines (5711, 6318, and 6319~\AA) were measured with the ROSA
package \citep{gra88}. The abundances were derived using the atmospheric
parameters determined for each star, interpolating within the \citet{kur93}
solar-scaled grid, with the overshooting option switched off. The adopted line lists and
atomic parameters are from \citet{gra03}.

Derived LTE abundances of Mg and Al are listed in Table 1
together with optical $B,V$,and $K$ magnitudes. For completeness, we
also list the atmospheric parameters derived in the full abundance analysis
(Carretta et al., in preparation). Adopted solar reference abundances are
7.43 dex for Mg and 6.23 dex for Al \citep{gra03,car04b}.

Typical star-to-star errors in abundance ratios are 0.058 and 0.047 dex for
[Al/Fe] and [Mg/Fe], respectively, due to internal errors in the adopted
atmospheric parameters (5 K in T$_{\rm eff}$, 0.04 dex in $\log g$, 0.03 dex in
]Fe/H], 0.05 kms$^{-1}$ in $v_t$) and $EW$ measurements; all errors were
estimated as described in details in \cite{car07b,car09b}.

\section{Three distinct groups on the RGB in NGC~2808}

\begin{figure} 
\includegraphics[bb=19 146 438 707, clip, scale=0.80]{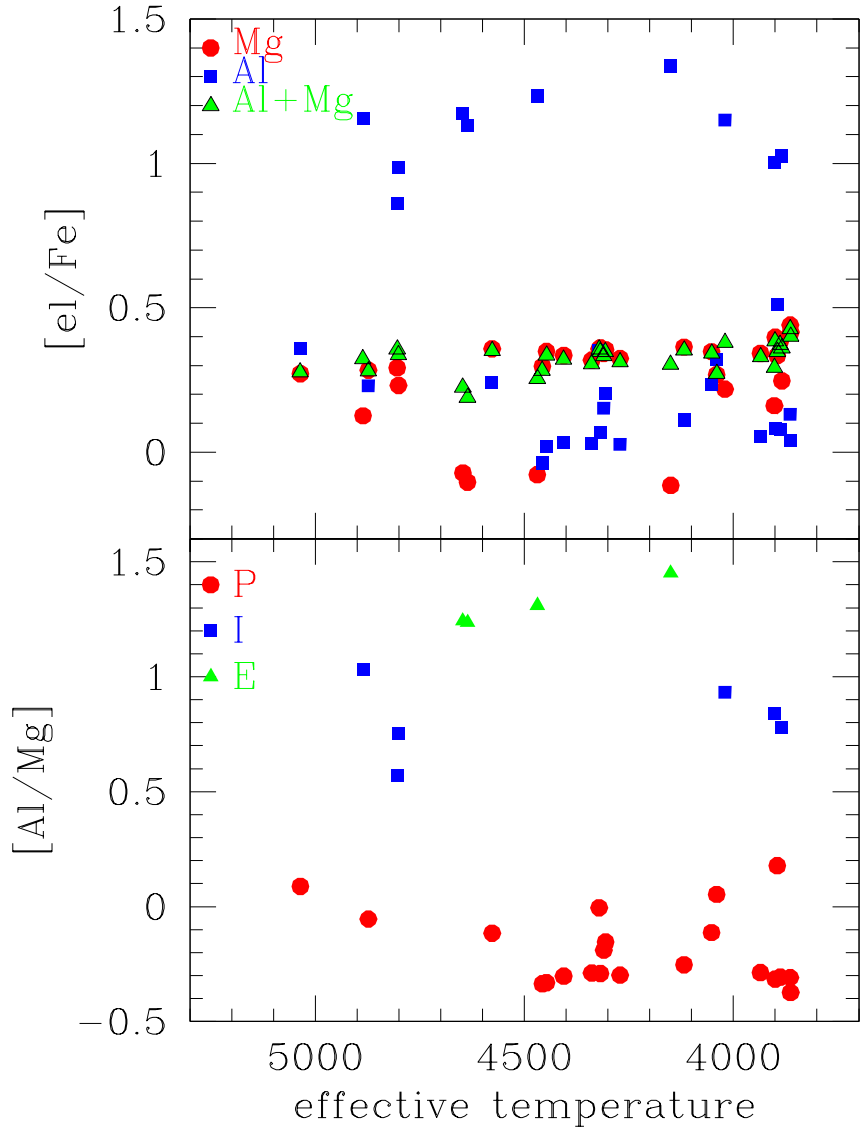}
\caption{Upper panel: [Mg/Fe] (circles), [Al/Fe] (squares), and 
[(Al+Mg)/Fe] (triangles) abundance ratios as a function of the effective
temperature. Lower panel: [Al/Mg] ratios as a function of the temperature.
Different symbols indicate stars of the three groups 
(see text and Fig.~\ref{f:fig3}).}
\label{f:fig2}
\end{figure}

The [Mg/Fe] and [Al/Fe] abundance ratios (and their sum) are plotted as a function
of the temperature in Fig.~\ref{f:fig2} (upper panel). No trend as a function of
T$_{\rm eff}$ (hence of the luminosity along the RGB) is discernible. Giants in
NGC~2808 show from moderate to large star-to-star abundance variations in Mg
content, with an average value [Mg/Fe]=0.26 dex and a rms scatter of
$\sigma=0.16$ dex (31 stars). The Al abundance presents much larger variations,
with a mean ratio [Al/Fe]=+0.46 dex and a higher rms scatter  ($\sigma=0.47$
dex).  On the other hand, the sum Al+Mg appears constant,  with an average
value [(Mg+Al)/Fe=]+0.33 and a very small dispersion ($\sigma=0.05$ dex).

We recovered all stars with a sub-solar Mg abundance found in the previous
studies, three from the NAO sample \citep{car09b} and one from the SV
sample (Paper I). All of them are actual detections, not upper limits.
The combined sample brings to four the number of stars with subsolar Mg
abundance in NGC~2808. Stars with such a low abundance are a notable exception
among the GCs, apart from NGC~2419 
\citep{muc12,coh12} and M~13 \citep{sne04}. The relatively high frequency of stars with strong
depletion in Mg, coupled to the high temperature ($\sim70$ MK) required for the
reactions of the Mg-Al cycle to occur, indicate that a fraction of stars in
NGC~2808 was formed by matter extremely processed by hot H-burning.

Even more interestingly, the better statistics from the present homogeneous
analysis clearly reveals that the RGB stars of the merged sample are not
continuously  distributed along the Mg-Al anticorrelation, but are instead
clustered into three distinct groups (Fig.~\ref{f:fig3}, upper panel). Using the
distribution of stars along the Na-O anticorrelation in several GCs,
\cite{car09a} divided the stellar population in each cluster in primordial
(first generation) and intermediate and extreme components of second generation.
By analogy, we associate the three groups along  the Mg-Al anticorrelation in
NGC~2808  with the same P, I, E components in order of increasing Al abundance
(and decreasing Mg content). The groups include  $68\pm15$\%, $19\pm8$\%, and
$13\pm4$\% of stars in our sample, respectively,  where the associated
uncertainties are Poisson errors.
Within the quoted statistical uncertainties these fractions of stars derived from
the Mg-Al anticorrelation are in excellent agreement with those estimated in
\citet{car06} from the O-normal ($61\pm7$\%), O-poor ($22\pm4$\%) and
super O-poor ($17\pm4$\%) RGB stars in NGC~2808 observed with GIRAFFE. A more
accurate comparison must however await the reanalysis with the new, homogeneous
temperature scale of all stars observed with both UVES and GIRAFFE.

No spurious effect due to the analysis can be responsible for the segregation of
stars into these groups. Even corrections for departures from LTE can have only
a negligible impact on this result, as shown in Fig.~\ref{f:fig2} (lower panel)
where we plot the [Al/Mg] ratios as a function of T$_{\rm eff}$, indicating
stars of different groups with different symbols. Giants with very similar
atmospheric parameters, that would be equally  affected by possible NLTE
corrections and uncertainties in atmospheric parameters, are neatly segregated
into the three P, I, E groups.

We show in the upper panel of Fig.~\ref{f:fig4} the distribution of the [Al/Mg]
ratios, that actually maximize the signal along the anticorrelation and allows
us to nicely  trace the three distinct clumps of stars separated by gaps at
[Al/Mg]$\sim 0.5$  dex and [Al/Mg]$\sim 1.15$ dex.  The
three components are characterized by very different average values of the
[Al/Mg] ratio: $-0.191\pm0.035$ dex ($\sigma=0.160$ dex, 21 P stars), 
$+0.818\pm0.065$ dex ($\sigma=0.158$ dex, 6 I stars), and $+1.310\pm0.050$ dex
($\sigma=0.100$ dex, 4 E stars).

To evaluate how sound is this division, in the lower panel of Fig.~\ref{f:fig4}
we plot the cumulative distribution of [Al/Mg] ratios for each of the P,I,E
group. A Kolmogorov-Smirnov test always allows to reject the null hypothesis
that the three components are extracted from the same parent population, {\bf
with $>99.999\%$ confidence.}

There is also a hint that the average metallicity slightly increases with
the [Al/Mg] ratio, from [Fe/H]$=-1.136$ dex ($\sigma=0.032$ dex) for the P
group to [Fe/H]$=-1.120$ dex ($\sigma=0.020$ dex) and 
[Fe/H]$=-1.110$ dex ($\sigma=0.006$ dex) for the I and E components,
respectively. Although these values cannot be considered formally different with a high
level of confidence, this finding is in qualitatively agreement with the
prediction that more He-enrichment in more polluted stars
\citep[e.g.][]{dan02} would also increase the  strength of metallic lines
in stars with the same original metal abundance \citep{boh79},
confirming the result obtained from a larger sample by \citet{bra10}
in NGC~2808.

\section{Discussion and conclusions}

In the present study we were able to provide a chemical tagging of three 
distinct groups of stars along the Mg-Al anticorrelation on the RGB in NGC~2808,
using the largest homogeneous set of Mg and Al abundances for giants in this 
cluster. The separation among the three groups on the RGB is much clearer in
NGC~2808 than in NGC~6752 \citep{car12,mil13}. On the other hand, the three MSs
in NGC~6752 seem to stand out less clearly than in the benchmark of multiple MSs
represented by NGC~2808 \citep[see][]{mil12,mil13}. This is not unexpected,
because NGC~2808 is much more massive than NGC~6752 and it is currently well
assessed that  the cluster total mass is the main parameter driving the
phenomenon of multiple populations in GCs \citep{car10}.

The fractions of stars in each of the P, I, and E components we found on the RGB
are in very close agreement with the number ratios found for single MS stars in
NGC~2808 by \citet{mil12}: $62\pm2$\%, $24\pm2$\%, and $14\pm3$\% for
the red, middle and blue MS, respectively. 

\begin{figure} 
\includegraphics[scale=0.80]{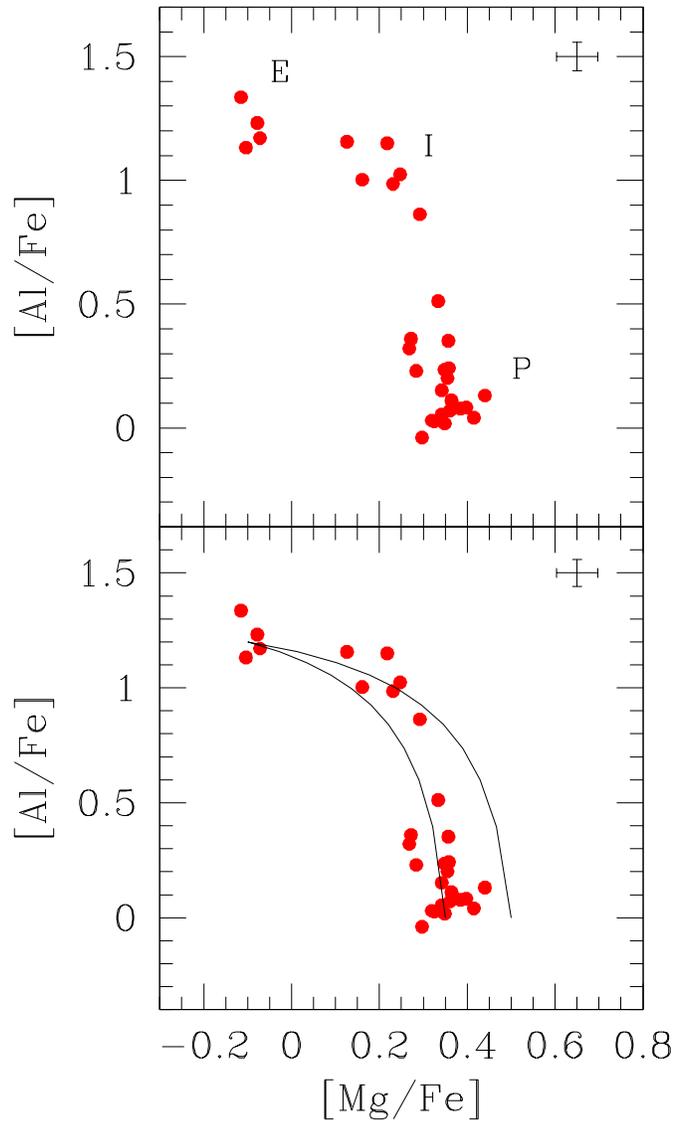}
\caption{Upper panel: Mg-Al anticorrelation in NGC~2808. Star to star error
bars are indicated. Lower panel: the same plot, with two dilution
models superimposed, starting at different primordial Mg levels.}
\label{f:fig3}
\end{figure}

\begin{figure} 
\includegraphics[bb=124 150 420 708,scale=0.80]{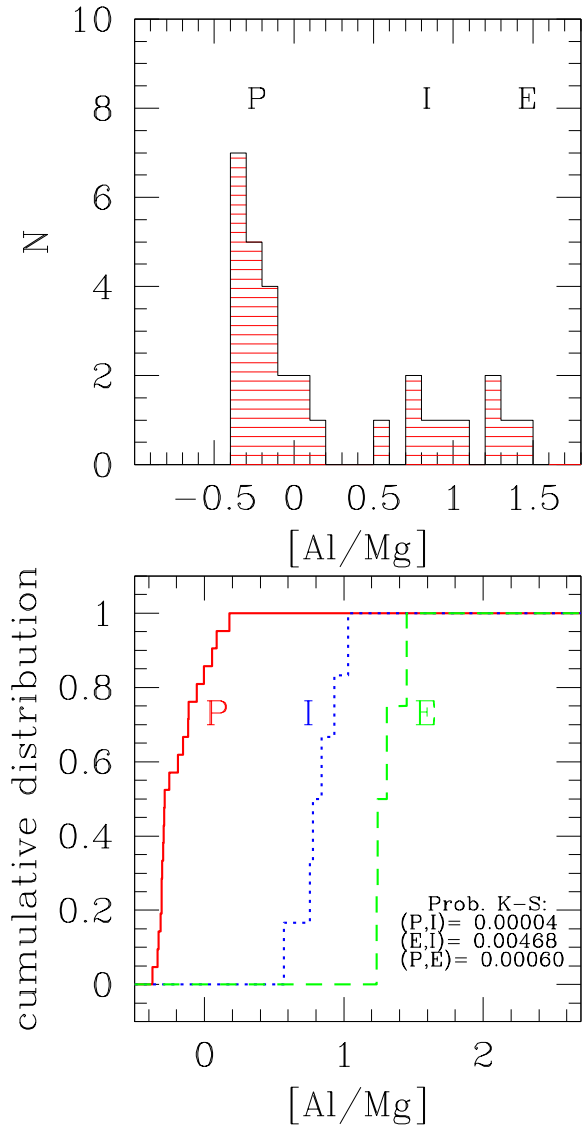}
\caption{Upper panel: distribution of [Al/Mg] ratios for the entire sample of
giants in NGC~2808. Lower panel: cumulative distributions of [Al/Mg] ratios in
the P (solid line), I (dotted line), and E (dashed line) groups.
The probabilities of the Kolmogorov-Smirnov test are also indicated.}
\label{f:fig4}
\end{figure}

Having found three discrete populations of RGB stars, each characterized by a
different chemical composition, we may ask what is the impact of this result on
the formation scenario and the elusive nature of first generation polluters.
Two commonly adopted scenarios (where the polluters are either rotating massive stars,
\citealt{dec07} or intermediate-mass AGB stars, \citealt{ven01}) 
share the need of dilution of the nuclearly processed matter with unprocessed,
pristine gas \citep{pra07,der11}. We then plot in the
lower panel of Fig.~\ref{f:fig3} two simple dilution models, as in 
\citet{car09b}. The first, adopting as polluted and original abundance ratios the
extremes observed in NGC~2808, is able to reproduce the E and P groups: however,
the I component is clearly left out. To be matched by this model, [Mg/Fe] and
[Al/Fe] should be lowered, on average, by about 0.10 and 0.20 dex, respectively:
the last value is about 4$\sigma$ off the plausible range.
On the other hand, if we arbitrarily change the starting values of Mg and Al
to fit simultaneously the I and E components, the primordial level of
Mg and Al would be inconsistent with what is actually observed.
To summarize, there is no way to reproduce the intermediate component by
mixing primordial and extremely polluted gas. As in the previous case of
NGC~6752 we must conclude again that different polluters acted to
produce the chemical composition observed in the I and E groups.

The strongly Mg-depleted stars observed in NGC~2808 represent a difficult
problem for aforementioned formation scenarios. Discrete distribution of
He values has been invoked as a likely way to explain both the three distinct
MSs and the multimodal distribution of stars on the HB of NGC~2808
\citep{dan05}. The bluest MS is reproduced using a model with a helium mass
fraction of $Y\sim0.38$, and this should then be the He content of stars in the
Mg-poor group, the E component. However, in one of the latest formulation of the
AGB-based formation scenario \citep{der12} the largest Mg depletions are not
from the super AGB, but come from AGB stars of 5-6$M_\odot$. The He content of
these stars is $Y\leq0.34$  \citep[see Tab. 1 in][]{der12}, which is the
threshold above which the  authors adopt a deep-mixing scheme able to decrease
the O abundance in second generation He-rich  stars down to the very low values
[O/Fe]$\sim -0.8 \div -1.0$ observed in GCs like NGC~2808 (Paper I) or M~13
\citep{joh12}. Unfortunately, the very Mg-poor stars of the E component are also
the most O-poor stars in NGC~2808 (Carretta et al., in preparation). Hence, as
advanced by \citet{der12}, this extreme population constitutes a problem for any
pollution  model and for the deep mixing scenario.  On the other hand, efficient
destruction of Mg and enhancement of Al are apparently produced by FRMS only at
the end of the MS, but with simultaneous  strong Na destruction \citep{dec07},
which is again not well matched by observations.

In summary:

1) we put on a better statistical base the known Mg-Al anticorrelation observed
in  NGC~2808 \citep{car09b}, more than doubling the size of the sample
of stars with homogeneous Mg and Al abundances.
The anticorrelation is very extended, reaching a regime where very Mg-poor
giants are found, a rare occurrence among GC stars.

2) for the first time we found clear evidence of three discrete populations each
with distinctly homogeneous chemical composition on the RGB of NGC~2808. The
first group P has the primordial composition of field stars of similar
metallicity, and the I and E components show signatures of increasing processing of
matter in hot H-burning.

3) the fractions of P,I,E stars along the Mg-Al anticorrelation are in excellent
agreement with the number ratios found by \citet{mil12} for the three MSs
in NGC~2808. We conclude that P, I, and E stars represent the progeny of the
red, intermediate and blue main sequences, with increasing He content.

4) No simple dilution model appears to be able to simultaneously reproduce the
chemistry of the three discrete components in NGC~2808, suggesting again that in
this cluster two different classes of first generation polluters were at work.
This is also seen in NGC~6752, another GCs with three discrete stellar
populations spectroscopically detected on the RGB.

Future steps include enlarging the set of Al abundances in this GC: a program
to measure the [Al/Fe] ratios of more than 100 giants in NGC~2808 using the HR21
setup of GIRAFFE and the strong Al doublet at 8772-74~\AA\ was just granted
observing time at the ESO@VLT.
Moreover, the homogeneous reanalysis of  other elements, in particular the
proton-capture species O, Na, Si, with the proper temperature scale, is in
progress, to provide the full network of correlations and anticorrelations in 
this peculiar globular cluster.

\acknowledgements
We thank Angela Bragaglia and Michele Bellazzini for valuable help and 
suggestions. This publication makes use of data products from the Two Micron All
Sky Survey, which is a joint project of the University of Massachusetts and the
Infrared Processing and Analysis Center/California Institute of Technology,
funded by the National Aeronautics and Space Administration and the National
Science Foundation.  This research has been funded by PRIN INAF 2011 "Multiple
populations in globular clusters: their role in the Galaxy assembly" (PI E.
Carretta), and PRIN MIUR 2010-2011, project ``The Chemical and Dynamical
Evolution of the Milky Way and Local Group Galaxies'' (PI F. Matteucci).

\clearpage
\begin{deluxetable}{rrrrrcccccc}
\tabletypesize{\scriptsize}
\tablecaption{Magnitudes, atmospheric parameters, [Mg/Fe], and [Al/Fe] ratios
for red giants in NGC~2808.\label{abu}}
\tablehead{\colhead{star}&
\colhead{RA,DEC}&
\colhead{$B$}&
\colhead{$V$}&
\colhead{$K$}&
\colhead{T$_{\rm eff}$}&
\colhead{$\log g$}&
\colhead{[A/H]}&
\colhead{$v_t$}&
\colhead{nr [Mg/Fe] $\sigma$}&
\colhead{nr [Al/Fe] $\sigma$}\\
}
\startdata
  8739 &9 11 51.20  -64 48 37.54& 15.761 & 14.288 & 10.693 & 4271 & 1.13 & -1.12 & 1.63 & 3  +0.325 0.122 & 2  +0.027 0.061  \\  
 38660 &9 12 39.87  -64 55 43.08& 15.672 & 14.290 & 10.875 & 4318 & 1.21 & -1.13 & 1.67 & 3  +0.361 0.070 & 2  +0.070 0.007  \\  
  8603 &9 12 14.05  -64 48 42.92& 15.847 & 14.432 & 10.957 & 4339 & 1.24 & -1.14 & 1.61 & 3  +0.319 0.088 & 2  +0.030 0.024  \\  
 10571 &9 12 41.12  -64 46 25.85& 15.806 & 14.376 & 10.844 & 4310 & 1.19 & -1.13 & 1.56 & 3  +0.342 0.115 & 2  +0.152 0.057  \\  
 30763 &9 11 31.62  -64 54 57.99& 15.927 & 14.606 & 11.465 & 4469 & 1.48 & -1.11 & 1.63 & 2~$-$0.078 0.020 & 2  +1.232 0.045  \\  
 49743 &9 12 36.80  -64 51 45.10& 16.353 & 15.168 & 12.159 & 4647 & 1.77 & -1.10 & 1.59 & 2~$-$0.072 0.045 & 2  +1.171 0.084  \\  
 34008 &9 11 27.52  -64 51 31.29& 16.346 & 15.119 & 12.115 & 4636 & 1.75 & -1.11 & 1.56 & 2~$-$0.104 0.022 & 2  +1.132 0.088  \\  
 38228 &9 12 30.97  -64 56  8.52& 16.211 & 14.981 & 11.887 & 4577 & 1.65 & -1.10 & 1.66 & 3  +0.358 0.022 & 2  +0.242 0.018  \\  
 10105 &9 12 21.15  -64 47 13.91& 15.976 & 14.669 & 11.417 & 4457 & 1.45 & -1.16 & 1.42 & 3  +0.297 0.104 & 2~$-$0.039 0.035  \\  
  7536 &9 12 31.71  -64 49 22.27& 15.812 & 14.372 & 10.829 & 4306 & 1.18 & -1.11 & 1.55 & 3  +0.355 0.095 & 2  +0.201 0.007  \\  
 44984 &9 12 45.88  -64 53  1.41& 15.855 & 14.535 & 11.219 & 4406 & 1.36 & -1.13 & 1.59 & 3  +0.335 0.066 & 2  +0.032 0.033  \\  
 33918 &9 11  1.69  -64 51 36.09& 15.776 & 14.331 & 10.887 & 4321 & 1.22 & -1.11 & 1.63 & 3  +0.357 0.088 & 2  +0.352 0.042  \\  
 48889 &9 12  8.51  -64 51 58.47& 15.139 & 13.341 &  9.253 & 3902 & 0.52 & -1.14 & 1.91 & 3  +0.161 0.060 & 2  +1.003 0.091  \\  
 51983 &9 12  2.50  -64 51 10.07& 15.328 & 13.474 &  9.182 & 3884 & 0.48 & -1.15 & 1.86 & 3  +0.247 0.080 & 2  +1.024 0.101  \\  
 47606 &9 12  6.66  -64 52 18.23& 15.257 & 13.436 &  9.103 & 3864 & 0.44 & -1.12 & 1.72 & 3  +0.440 0.052 & 2  +0.131 0.051  \\  
 51499 &9 12  7.35  -64 51 17.80& 15.157 & 13.435 &  9.382 & 3935 & 0.57 & -1.21 & 1.83 & 3  +0.342 0.052 & 2  +0.055 0.049  \\  
 46580 &9 11 56.19  -64 52 35.39& 15.292 & 13.690 &  9.790 & 4040 & 0.78 & -1.13 & 1.81 & 3  +0.268 0.045 & 2  +0.321 0.008  \\  
 48609 &9 12 16.64  -64 52  2.92& 15.285 & 13.415 &  9.101 & 3863 & 0.44 & -1.14 & 1.84 & 3  +0.415 0.083 & 2  +0.041 0.047  \\  
 50761 &9 11 57.08  -64 51 29.68& 15.394 & 13.390 &  9.224 & 3895 & 0.31 & -1.17 & 1.65 & 3  +0.334 0.051 & 2  +0.512 0.006  \\  
 37872 &9 12 23.07  -64 56 34.30& 15.334 & 13.650 &  9.711 & 4020 & 0.71 & -1.10 & 1.84 & 3  +0.218 0.070 & 2  +1.150 0.045  \\  
 51454 &9 12  2.28  -64 51 18.51& 15.233 & 13.446 &  9.244 & 3900 & 0.51 & -1.18 & 1.78 & 3  +0.398 0.082 & 2  +0.083 0.023  \\  
 50119 &9 11 42.92  -64 51 39.75& 15.425 & 13.886 & 10.218 & 4150 & 0.93 & -1.11 & 1.80 & 2~$-$0.115 0.033 & 2  +1.336 0.035  \\  
 46422 &9 11 56.09  -64 52 37.90& 15.174 & 13.376 &  9.197 & 3888 & 0.44 & -1.18 & 1.93 & 3  +0.385 0.067 & 2  +0.078 0.007  \\  
 46099 &9 12 33.58  -64 52 43.09& 15.375 & 13.741 &  9.835 & 4052 & 0.76 & -1.15 & 1.85 & 3  +0.348 0.075 & 2  +0.235 0.006  \\  
 53390 &9 11 52.95  -64 50 47.86& 15.958 & 14.669 & 11.379 & 4447 & 1.43 & -1.12 & 1.49 & 3  +0.349 0.045 & 1  +0.018 9.999  \\  
 43217 &9 12 32.67  -64 53 32.87& 17.471 & 16.440 & 13.675 & 5036 & 2.41 & -1.11 & 1.59 & 1  +0.272 9.999 & 2  +0.360 0.011  \\  
 13983 &9 11 39.79  -64 48 25.46& 17.027 & 15.940 & 13.092 & 4886 & 2.16 & -1.11 & 1.10 & 2  +0.126 0.078 & 2  +1.156 0.011  \\  
 10201 &9 12 45.76  -64 47  5.39& 16.864 & 15.714 & 12.759 & 4801 & 2.02 & -1.11 & 1.43 & 2  +0.231 0.037 & 2  +0.986 0.021  \\  
 42886 &9 12 59.11  -64 53 38.61& 17.019 & 15.921 & 13.040 & 4873 & 2.14 & -1.14 & 0.64 & 1  +0.284 9.999 & 1  +0.230 9.999  \\  
 32685 &9 11 27.07  -64 52 44.38& 16.794 & 15.656 & 12.772 & 4804 & 2.03 & -1.11 & 0.61 & 2  +0.292 0.025 & 2  +0.863 0.006  \\  
 56032 &9 11 45.58  -64 50  4.21& 15.579 & 13.976 & 10.095 & 4118 & 0.87 & -1.07 & 1.78 & 3  +0.364 0.056 & 2  +0.111 0.002  \\  
\enddata
\tablenotetext{a}{Identification, coordinates, and $B,V$ magnitudes from Bedin et al. (2000)}
\tablenotetext{b}{$K$ magnitudes from Skrutskie et al. (2006)}
\tablenotetext{c}{Atmospheric parameters from Carretta et al. (in preparation).}
\tablenotetext{d}{$\sigma$ is the $rms$ scatter of the mean.}
\end{deluxetable}

\clearpage

\clearpage

\clearpage

\clearpage

\end{document}